\documentclass[12pt]{article}

\usepackage[margin=0.7in]{geometry}
\usepackage{graphicx}
\usepackage{authblk}
\usepackage{natbib}
\usepackage{hyperref}
\usepackage{setspace}
\usepackage{amsmath}
\usepackage{amssymb}
\usepackage{booktabs}
\usepackage{multirow}
\usepackage{threeparttable}
\usepackage{array}
\usepackage{tabularx}
\usepackage{makecell}
\usepackage{float}
\usepackage[tablesonly,nomarkers]{endfloat}

\AtBeginTables{\clearpage\section*{Tables}}
\usepackage{bm}
\usepackage{adjustbox}

\title{Using binary silver labels in electronic health records-based computable phenotyping algorithms}
\author[1,*]{Shuhe Wang}
\author[2]{Matthew T. Slaughter}
\author[3,1]{Jennifer C. Nelson}
\author[3,1,4]{Brian~D. Williamson}
\affil[1]{Department of Biostatistics, University of Washington}
\affil[2]{Kaiser Permanente Center for Health Research}
\affil[3]{Biostatistics Division, Kaiser Permanente Washington Health Research Institute}
\affil[4]{Vaccine and Infectious Disease Division, Fred Hutchinson Cancer Center}

\date{\today}

\begin{document}

\maketitle


\begin{abstract}
Gold-standard phenotype labels are often unavailable at scale in electronic health record (EHR) studies because they require manual review of clinical records. Weakly supervised phenotyping methods address this problem by fitting prediction algorithms using silver-standard labels, such as diagnosis-code counts or natural language processing (NLP) mention counts, together with routinely available structured features from claims or EHR and NLP-based features. PheNorm is a widely used weakly supervised method, but its original formulation was developed for count-valued silver labels and relies on log-transformation, utilization normalization, and Gaussian mixture modeling, using an expectation-maximization (EM) algorithm to obtain final predicted phenotype probabilities. These modeling steps are not directly suited to binary silver labels, even though binary indicators such as medication administration, laboratory values above a threshold, or clinically meaningful NLP indicators are common in EHR applications and may be highly predictive of true phenotype status. We propose Binary PheNorm, an extension of PheNorm that uses binary silver labels directly in the corruption-and-regression denoising step and produces a continuous phenotype score without requiring an EM calibration step. We also consider a lasso-regularized version that can be used for either binary or count silver labels in high-dimensional sparse EHR settings. In simulations, Binary PheNorm achieved strong discrimination using binary silver labels alone, and often improved performance when combined with count labels. In two applications, Binary PheNorm improved substantially over using the binary silver label itself as a classifier, without PheNorm denoising: for predicting anaphylaxis, the AUC increased from 0.793 for an epinephrine-mention indicator to 0.891--0.892 after Binary PheNorm; for predicting acute pancreatitis, the AUC increased from 0.736 for a lipase laboratory-threshold indicator to 0.805--0.819 after Binary PheNorm. These results support Binary PheNorm as a practical weakly supervised approach when gold-standard labels are limited and informative binary silver labels are available.
\end{abstract}

\doublespacing

\section{Introduction}

Electronic health record (EHR)-based computable phenotyping algorithms are increasingly used in clinical research and safety surveillance. For example, the United States Food and Drug Association (FDA) Sentinel Initiative and the Vaccine Safety Datalink use routinely collected healthcare data to support postmarket safety surveillance of medical products \citep{platt2018fda,platt2012us, baggs2011vaccine, chen1997vaccine, mcneil2014vaccine}. A central challenge in EHR-based phenotyping is that gold-standard phenotype labels usually require manual chart review, which is expensive and difficult to obtain at scale \citep{yu2018enabling}. This challenge is especially important for clinically complex outcomes, such as anaphylaxis and acute pancreatitis, where diagnosis codes alone do not reliably identify true cases \citep{walsh2013validation,carrell2023improving,floyd2023validation}. Weakly supervised phenotyping methods address this problem by using silver-standard labels, which are imperfect but informative proxies for the unobserved true phenotype \citep{yu2018enabling, bach2017learning, ahuja2020surelda, liao2019high}. These silver labels may be based on diagnosis codes, natural language processing (NLP) mentions, medication records, laboratory values, or other EHR-derived variables. Although some processing may be needed to construct them, silver labels can often be generated through scalable and largely automated EHR or NLP pipelines. As a result, they can be available for model training in much larger cohorts than chart-reviewed gold-standard labels, while gold-standard labels can be reserved for performance evaluation.

Several weakly supervised methods have been developed to reduce reliance on manually labeled training data. PheNorm is one widely used example \citep{yu2018enabling}. It was originally developed for count-valued silver labels, such as diagnosis-code counts or NLP mention counts, and combines these silver labels with auxiliary structured and NLP-derived EHR features to construct predicted phenotype probabilities \citep{yu2018enabling}. Related methods, including MAP and sureLDA, have also been proposed for weakly supervised EHR phenotyping \citep{liao2019high,ahuja2020surelda}. However, these methods do not directly address the setting in which the main available silver labels are binary indicators. MAP also models aggregated ICD and NLP count features using latent mixture models, and sureLDA uses PheNorm or MAP-based probabilities to initialize its downstream topic model. Thus, although these methods extend weakly supervised phenotyping in other directions, they do not provide a binary-label-specific denoising procedure for medication indicators, laboratory-threshold indicators, or clinically meaningful NLP indicators.

Binary silver labels are common in clinical applications and may contain important phenotype information. For example, an indicator of epinephrine administration or an epinephrine mention may be informative for anaphylaxis phenotyping because epinephrine is the first-line treatment for anaphylaxis \citep{cardona2020world,golden2024anaphylaxis}. Similarly, an indicator that a patient's maximum lipase level exceeds three times the laboratory-specific upper limit of normal may be informative for acute pancreatitis phenotyping because this threshold is commonly used in acute pancreatitis case definitions \citep{floyd2023validation,bann2026comparison}. Such labels are not counts, and treating them as count-valued inputs in the Original PheNorm framework is not well motivated. In addition, binary silver labels are not directly supported by the current Original PheNorm software implementation, which was designed for count-valued silver labels and does not accommodate binary silver labels as primary surrogate inputs. 

In this paper, we propose Binary PheNorm, an extension of PheNorm that uses binary silver labels directly in the corruption-and-regression denoising step. The proposed approach preserves the weakly supervised structure of PheNorm: gold-standard labels are not used for model training, but are used for evaluation when available. Binary PheNorm also avoids the mixture-normal assumption and EM calibration step required by Original PheNorm. We consider combined models that use both binary and count silver labels, as well as a lasso-regularized version for sparse, high-dimensional EHR and NLP feature matrices. Finally, we discuss how to incorporate binary silver labels in sureLDA procedure. We evaluate the proposed approach in simulations and in two EHR phenotyping applications, anaphylaxis and acute pancreatitis.

The remainder of this manuscript is organized as follows. In Section~\ref{sec:methods}, we review Original PheNorm and describe Binary PheNorm, combined binary-count models, lasso-regularized denoising, and the evaluation strategy. In Section~\ref{sec:simulations}, we present the simulation studies. In Section~\ref{sec:real_data_analysis}, we evaluate the proposed approach in two EHR phenotyping applications, anaphylaxis and acute pancreatitis. We conclude with a discussion of the implications, limitations, and appropriate use of the method in Section~\ref{sec:discussion}.

\section{Methods}\label{sec:methods}

\subsection{Notation and target problem}

Throughout, scalar variables are denoted by ordinary italic lowercase letters, subject-level vectors by bold italic lowercase letters, and data matrices by bold upright uppercase letters. Let \(i=1,\ldots,n\) index subjects or encounters. Let \(y_i\in\{0,1\}\) denote the true phenotype. In the weakly supervised setting we consider, \(y_i\) is not used for model fitting and may be unavailable for many records. The available information consists of silver labels and EHR covariates.

Let \(\bm{b}_i=(b_{i1},\ldots,b_{ik_b})^\top\) denote binary silver labels, where each \(b_{ij}\in\{0,1\}\). Let \(\bm{c}_i=(c_{i1},\ldots,c_{ik_c})^\top\) denote count-valued silver labels, and let \(\bm{z}_i=(z_{i1},\ldots,z_{im})^\top\) denote auxiliary EHR covariates, which can consist of structured covariates and covariates obtained through NLP of clinical text. The corresponding data matrices are denoted by \(\mathbf{B}\in\{0,1\}^{n\times k_b}\), \(\mathbf{C}\in\mathbb{R}^{n\times k_c}\), and \(\mathbf{Z}\in\mathbb{R}^{n\times m}\). The goal is to construct a continuous phenotype score \(t_i\) using \(\mathbf{B}\), \(\mathbf{C}\), and \(\mathbf{Z}\), without using the gold-standard labels \(y_i\) for training. When a chart-reviewed subset is available, the gold-standard labels are used only to evaluate discrimination and threshold-based classification performance.

\subsection{Original PheNorm for count and continuous silver labels}

PheNorm was developed for count-based silver labels, such as diagnosis-code counts or NLP mention counts \citep{yu2018enabling}. For a count silver label \(c_{ij}\), the original algorithm first applies a log transformation and adjusts for healthcare utilization. Let \(n_i\) denote a utilization measure, such as the total number of notes or encounters. The normalized count silver label is
\[
\widetilde c_{ij}
=
\log(1+c_{ij})-\alpha_j\log(1+n_i),
\]
where \(\alpha_j\) is chosen using the mixture-model criterion described in the Original PheNorm paper \citep{yu2018enabling}.

After normalization, PheNorm combines the normalized count silver labels with auxiliary EHR covariates. Let
\[
\bm{x}_i=(\widetilde{\bm{c}}_i^\top,\bm{z}_i^\top)^\top,
\qquad
\text{and}
\qquad
\widetilde{\bm{c}}_i=(\widetilde c_{i1},\ldots,\widetilde c_{ik_c})^\top .
\]
The algorithm then uses a corruption-and-regression denoising step. A corrupted version of \(\bm{x}_i\), denoted by \(\bm{x}_i^{(c)}\), is created by randomly replacing some feature values with their column means. For each count silver label \(j\), PheNorm fits
\[
\widetilde c_{ij}
=
\beta_{0j}+\bigl(\bm{x}_i^{(c)}\bigr)^\top\bm{\beta}_j+\varepsilon_{ij},
\qquad
\text{for}\qquad i=1,\ldots,n .
\]
and evaluates the fitted regression at the uncorrupted feature vector to obtain a denoised score,
\[
t_{ij}
=
\widehat\beta_{0j}+\bm{x}_i^\top\widehat{\bm{\beta}}_j .
\]

In Original PheNorm, the denoised count-label score is further mapped to an estimated phenotype probability using a two-component mixture-model calibration step \citep{yu2018enabling}. This final step relies on the assumption that the transformed count-label score is well approximated by the specified mixture distribution.

\subsection{Binary PheNorm}

Binary PheNorm is designed for settings in which the silver labels used for denoising are binary. The method does not transform the binary label and does not impose a Gaussian mixture model on it. For a single binary silver label, the marginal distribution only identifies the overall event probability, so different combinations of latent class proportions and class-specific Bernoulli probabilities can yield the same observed distribution. More generally, finite mixtures of multivariate Bernoulli distributions are not identifiable \citep{gyllenberg1994non}. Therefore, applying an EM mixture model to binary silver labels may produce component parameters and posterior probabilities that are not uniquely determined by the observed data. Binary PheNorm instead uses the binary silver label directly as the target in the PheNorm denoising regression.

Consider first the setting in which the available silver labels are binary, together with auxiliary EHR covariates. Define \(\bm{x}_i=(\bm{b}_i^\top,\bm{z}_i^\top)^\top\in\mathbb{R}^{k_b+m}\), and let \(\mathbf{X}\in\mathbb{R}^{n\times(k_b+m)}\) be the corresponding feature matrix. As in PheNorm, construct a corrupted vector \(\bm{x}_i^{(c)}\) by replacing randomly selected entries with column means. For each binary silver label \(j\), fit the linear denoising regression
\[
b_{ij}
=
\beta_{0j}+\bigl(\bm{x}_i^{(c)}\bigr)^\top\bm{\beta}_j+\varepsilon_{ij},
\qquad
\text{for}\qquad i=1,\ldots,n .
\]
and define \(t_{ij}=\widehat\beta_{0j}+\bm{x}_i^\top\widehat{\bm{\beta}}_j\). When several binary silver labels are available, the label-specific scores can be averaged as \(t_i=k_b^{-1}\sum_{j=1}^{k_b}t_{ij}\), or combined using another prespecified rule, as in Original PheNorm. In this paper we use simple averaging to preserve the weakly supervised character of the method and avoid estimating combination weights from gold-standard labels.

The choice of linear regression for a binary target is intentional. Logistic regression is a natural model for a binary response, but in this denoising setting the target is a silver label rather than the true phenotype. The purpose of the regression is to extract a stable continuous score from shared information in the silver labels and covariates, not to estimate the conditional probability of the true phenotype. In our numerical experiments, logistic regression was less stable when binary silver labels were sparse or highly predictive. This behavior is consistent with the known separation problem in logistic regression, in which highly predictive or unbalanced covariates can lead to divergent maximum likelihood estimates \citep{heinze2002solution}. Ordinary least squares gives a convex and numerically simple denoising step and performed better empirically in the settings considered here. We provide a supplementary comparison of logistic-regression and linear-regression denoising in Supplementary Appendix Section~S1.

The output of Binary PheNorm should be interpreted as a phenotype score rather than a calibrated phenotype probability. This distinction is important because the fitted values from the linear denoising regression are not constrained to lie in the interval \([0,1]\). If a probability-scale output is desired, the score can be truncated to \([0,1]\), or mapped to \((0,1)\) using an inverse-logit transformation. However, such transformations should be viewed as post-processing steps rather than evidence that the score is calibrated. By avoiding the mixture-model calibration step, Binary PheNorm makes fewer distributional assumptions than Original PheNorm, but it also does not claim probability calibration without additional validation. For many phenotyping tasks, a continuous score is sufficient for ranking patients, enriching a chart-review sample, or applying a prespecified classification threshold.

\subsection{Combining Original PheNorm and Binary PheNorm within the PheNorm framework}
\label{subsec:phenorm_framework}

Original PheNorm and Binary PheNorm can be viewed as two implementations of the same broader PheNorm framework. Both use silver labels as noisy surrogates for the unobserved phenotype, borrow information from auxiliary EHR and NLP-derived covariates through the corruption-and-regression denoising step, and do not use gold-standard labels for model fitting. They differ mainly in the type of silver label being used and the algorithm used to obtain the score. Original PheNorm is used for count or continuous silver labels after the appropriate transformation and utilization normalization, whereas Binary PheNorm uses binary silver labels directly on their original scale.

When both binary and count silver labels are available, the two approaches can be combined within this common framework. In the combined setting, which we call Combined PheNorm, count labels are first transformed and utilization-normalized as in Original PheNorm, while binary labels are left unchanged as in Binary PheNorm. The combined feature vector is
\[
\bm{x}_i=(\bm{b}_i^\top,\widetilde{\bm{c}}_i^\top,\bm{z}_i^\top)^\top.
\]
The same corruption-and-regression denoising step is then applied using the available silver labels as regression targets. Binary silver labels contribute Binary PheNorm scores, and transformed count silver labels contribute Original PheNorm scores. The resulting label-specific scores are averaged over the silver labels included in the fitted model. This combination is still weakly supervised, because no gold-standard labels are used to estimate the score or to choose weights across silver labels.

This Combined PheNorm model is useful when binary and count silver labels contain complementary information about the latent phenotype. For example, a count label may capture repeated documentation of a condition, whereas a binary medication or laboratory indicator may capture a clinically specific event. If both types of silver labels are informative and not redundant, combining Original PheNorm and Binary PheNorm can improve discrimination. We evaluate this Combined PheNorm framework in both simulations and real-data analyses.

\subsection{Use within sureLDA}

Binary PheNorm and Combined PheNorm can also be incorporated into downstream weakly supervised phenotyping algorithms that use PheNorm for initialization. For example, sureLDA first uses PheNorm to obtain initial phenotype probabilities from silver-standard surrogates and healthcare utilization, and then uses these probabilities as Dirichlet hyperparameters in a guided latent Dirichlet allocation model \citep{ahuja2020surelda}. When binary silver labels are available, this initialization step can be replaced by Binary PheNorm; when both binary and count silver labels are available, it can be replaced by Combined PheNorm.

Because Binary PheNorm produces a continuous phenotype score rather than a calibrated probability, the score can be mapped to a probability scale before being used in sureLDA, as discussed above. The later sureLDA steps can then be left unchanged. This provides a direct way to extend sureLDA to settings where binary silver labels are informative but Original PheNorm is not directly applicable.

\subsection{Lasso-regularized denoising}

EHR phenotyping often involves many structured and NLP-derived predictors. Many of these variables are sparse, weakly informative, or redundant. In such settings, the unregularized denoising regression may be unstable. We observed this in the anaphylaxis data example described more fully below, where over repeated applications of PheNorm with different random number seeds, the algorithm failed to converge 30\% of the time. We therefore consider a lasso-regularized version of both Binary PheNorm and Original PheNorm \citep{tibshirani1996regression}.

For a binary target \(b_{ij}\), with corrupted matrix \(\bm{x}_i^{(c)}\), the lasso denoising estimator is
\[
\begin{aligned}
(\widehat\beta_{0j},\widehat{\bm{\beta}}_j)
=
\arg\min_{\beta_0,\bm{\beta}}
\bigg\{
\frac{1}{n}\sum_{i=1}^n
\left(
 b_{ij}-\beta_0-\bigl(\bm{x}_i^{(c)}\bigr)^\top\bm{\beta}
\right)^2
+
\lambda_j\|\bm{\beta}\|_1
\bigg\}.
\end{aligned}
\]
For a normalized count target \(\widetilde c_{ij}\), the same objective is used with \(b_{ij}\) replaced by \(\widetilde c_{ij}\). The score is again computed on the uncorrupted feature vector:
\[
t_{ij}=\widehat\beta_{0j}+\bm{x}_i^\top\widehat{\bm{\beta}}_j.
\]
The penalty parameter \(\lambda_j\) is selected by cross-validation within the silver-label denoising regression, implemented using standard coordinate-descent algorithms for penalized regression \citep{friedman2010regularization}. This cross-validation uses the silver label as the regression target and does not use the gold-standard phenotype label.

\subsection{Model fitting and evaluation}

The defining feature of Original PheNorm and Binary PheNorm is that the gold-standard phenotype labels are not used for estimating the score. This affects how training and evaluation should be interpreted. In supervised learning, train/test splitting or \(K\)-fold cross-validation is commonly used to estimate out-of-sample predictive performance and to reduce optimism from evaluating a model on the same gold-standard outcome used for training \citep{hastie2009elements,arlot2010survey}.In weakly supervised PheNorm, the model is fit using only silver labels and covariates; therefore, the usual supervised-learning concern is different. Gold-standard labels are still needed to estimate performance, but they are not used to estimate the PheNorm score.

Fitting a final weakly supervised model on the full available EHR cohort is natural when the goal is deployment or cohort construction, because using more silver-label and covariate information can improve the stability of the denoising model. Because gold-standard labels are not used in model fitting, we can evaluate the resulting score on the gold-labeled data without data splitting. Nevertheless, for transparent reporting, we compare three evaluation strategies in simulations and the real-data analyses, as described in Sections~\ref{subsec:sim_data_splitting} and~\ref{sec:real_data_analysis}: a no-split analysis, a train/test split, and cross-validation. The no-split analysis estimates the score using all available records and evaluates performance in the chart-reviewed sample. The train/test and cross-validation analyses train the algorithm on separate data from the evaluation dataset, and evaluate whether it is necessary to correct for overfitting when developing and evaluating a score. Similar results across these strategies would support the interpretation that the method is not overfitting.

Performance is summarized using the area under the receiver operating characteristic curve (AUC) and common threshold-based metrics in binary classification tasks \citep{hanley1982meaning}. AUC is used as a threshold-free measure of discrimination. Sensitivity, specificity, positive predictive value (PPV), negative predictive value (NPV), and accuracy are reported at prespecified score thresholds. In the real-data analyses described in Section~\ref{sec:real_data_analysis}, sampling weights are used when evaluating performance because the chart-reviewed records were obtained from sampling designs rather than simple random samples of all eligible events.

\section{Numerical experiments}\label{sec:simulations}

\subsection{Objectives}

Our numerical experiments were designed to evaluate three main aspects of the proposed method. First, we compared no-split, train/test split, and cross-validation strategies in simulation to assess whether performance estimates were sensitive to the evaluation design, given that gold-standard labels were not used to estimate the weakly supervised PheNorm scores. Second, we evaluated whether Binary PheNorm could recover phenotype signal when the available silver labels were binary and the model also included auxiliary EHR covariates. We also examined whether performance improved when more than one binary silver label was available, whether binary silver labels could complement count silver labels within the broader PheNorm framework, and how performance changed in rare-outcome settings. Third, we evaluated whether lasso regularization improved performance when the covariate matrix was high-dimensional and sparse.

\subsection{Investigating the need for data splitting in PheNorm}\label{subsec:sim_data_splitting}

We first conducted a simulation to examine whether performance estimates were sensitive to the evaluation design. Each Monte Carlo replication generated \(n=1000\) subjects. The true phenotype was generated from \(y_i\sim\mathrm{Bernoulli}(0.5)\). We generated \(d=10\) covariates from phenotype-specific normal distributions: \(z_{ij}\mid y_i=1\sim N(0.7,0.5^2)\) and \(z_{ij}\mid y_i=0\sim N(0.3,0.5^2)\), for \(j=1,\ldots,10\). Let \(\mathbf{Z}\in\mathbb{R}^{n\times d}\) denote the resulting covariate matrix.

One binary silver label was generated by perturbing the true phenotype to create errors in the binary label. Specifically, the binary silver label \(b_i\) satisfied \(P(b_i=y_i)=0.8\). We used two-sided misclassification, under which a proportion \(0.2\) of labels was selected at random and flipped. This setting represents a non-rare outcome setting with one noisy but informative binary silver label.

In each replication, 20\% of subjects were randomly designated as the gold-standard validation set, while the remaining 80\% were treated as non-gold subjects. We compared three evaluation strategies. In the no-split analysis, the weakly supervised model was fit using all subjects and evaluated only on the gold-standard validation set. In the train/test split analysis, the model was fit using the non-gold subjects and evaluated on the gold-standard validation set. In the 10-fold cross-validation analysis, the cross-validation folds were formed only within the gold-standard validation set. For each fold, one-tenth of the gold-standard validation set was held out for evaluation. The model was fit using all non-gold subjects together with the remaining nine-tenths of the gold-standard validation set. Thus, the non-gold subjects were included in the training data for every cross-validation fold, and the held-out fold always came only from the gold-standard validation set.

In all three strategies, the true phenotype labels were used only for performance evaluation. They were not used to fit the PheNorm score, even when gold-set subjects were included in the training data for the no-split and cross-validation analyses. This evaluation-strategy simulation was repeated over 1000 independent Monte Carlo replications. Performance was summarized by AUC, sensitivity, specificity, PPV, and accuracy. Threshold-based metrics were computed using the median score threshold.

\begin{table}[htbp]
\centering
\small
\begin{threeparttable}
\caption{Simulation performance across evaluation strategies.}
\label{tab:sim_split_comparison}

\setlength{\tabcolsep}{4pt}
\renewcommand{\arraystretch}{1.15}

\begin{tabular}{@{}lccccc@{}}
\toprule
Evaluation strategy & AUC & Sens. & Spec. & PPV & Acc. \\
\midrule
No split
& 0.907 (0.025) & 0.809 (0.038) & 0.808 (0.037) & 0.807 (0.039) & 0.808 (0.026) \\
Train/test split
& 0.902 (0.024) & 0.810 (0.044) & 0.807 (0.042) & 0.808 (0.039) & 0.808 (0.025) \\
10-fold CV
& 0.906 (0.022) & 0.809 (0.038) & 0.807 (0.038) & 0.815 (0.039) & 0.807 (0.025) \\
\bottomrule
\end{tabular}

\begin{tablenotes}[flushleft]
\footnotesize
\item[] Values are reported as mean (Monte Carlo standard deviation) over simulation replications. The simulation used the non-rare, two-sided misclassification setting with one binary silver label, 10 auxiliary covariates, and a randomly selected gold-standard validation set containing 20\% of subjects. Sensitivity, specificity, PPV, and accuracy were evaluated using the median score threshold. Sens. = sensitivity; Spec. = specificity; PPV = positive predictive value; Acc. = accuracy; CV = cross-validation.
\end{tablenotes}
\end{threeparttable}
\end{table}

The three evaluation strategies gave very similar results (Table~\ref{tab:sim_split_comparison}). Mean AUC was 0.907 for no split, 0.906 for 10-fold cross-validation, and 0.902 for the train/test split. Sensitivity, specificity, PPV, and accuracy were also nearly identical across strategies. These results suggest that, in this weakly supervised setting, performance estimates were not highly sensitive to whether the score was evaluated using no split, a train/test split, or cross-validation. This is consistent with the fact that the gold-standard phenotype labels were not used to fit the PheNorm score.

\subsection{Testing Binary PheNorm in main simulation scenarios}\label{subsec:sim_main}

Let \(y_i\) denote the true phenotype for subject \(i\). Each Monte Carlo replication generated \(n=1000\) subjects. In the main simulations, $y_i\sim\mathrm{Bernoulli}(p)$, where $p=0.5$ was used for non-rare outcome settings and $p=0.05$ was used as an illustrative rare-outcome setting. This rare-outcome setting was motivated by low-prevalence EHR phenotyping applications, where target phenotypes can occur in only a small fraction of records; for example, a recent EHR phenotyping study evaluated neonatal culture-negative sepsis with approximately 3\% prevalence \citep{kauffman2025infehr}. A healthcare utilization variable was generated as $n_i\sim\mathrm{Poisson}(2)+1$.

For a count silver label \(c_{ij}\), we first generated \(e_{ij}\mid y_i=y\sim N(\mu_y,\sigma_y^2)\), and then set
\[
c_{ij}=\log\left\lceil\exp(e_{ij})n_i^{1/2}\right\rceil .
\]
The main simulations used \(\mu_0=0.5\), \(\mu_1=1.0\), and \(\sigma_0=\sigma_1=0.4\). This count-label generating mechanism is favorable to Original PheNorm, because the transformed count label is generated from a phenotype-specific normal distribution, matching the type of distributional structure assumed by the original method.

We generated \(d=10\) covariates from phenotype-specific normal distributions: \(z_{ij}\mid y_i=1\sim N(0.7,0.5^2)\) and \(z_{ij}\mid y_i=0\sim N(0.3,0.5^2)\), for \(j=1,\ldots,10\). Let \(\mathbf{Z}\in\mathbb{R}^{n\times d}\) denote the resulting covariate matrix.

Binary silver labels were generated by perturbing the true phenotype to create errors in the binary label. For a binary silver label \(b_{ij}\), the main simulations used \(P(b_{ij}=y_i)=\rho\), with \(\rho=0.8\). We considered both one-sided and two-sided misclassification. Under two-sided misclassification, a proportion \(1-\rho\) of labels was selected at random and flipped. Under one-sided misclassification, errors were introduced in only one direction. The four main scenarios were: one-sided misclassification, two-sided misclassification, one-sided misclassification with a rare outcome, and two-sided misclassification with a rare outcome.

Within each scenario, we compared four silver-label configurations: one binary label, two binary labels, one count label, and one binary plus one count label. For each fitted model, the score was obtained using the same random-corruption denoising structure. Let \(\mathbf{S}_{\mathrm{fit}}\) denote the silver-label matrix used in a given fitted model and define
\[
\mathbf{X}_{\mathrm{fit}}=[\mathbf{S}_{\mathrm{fit}},\mathbf{Z}].
\]
The denoising regression was fit using the corrupted version of \(\mathbf{X}_{\mathrm{fit}}\), and the fitted regression was evaluated on the uncorrupted matrix to obtain the final score, as described in Section~2.4.

Each main simulation scenario and silver-label configuration was repeated by generating data, fitting PheNorm models, and evaluating performance over 2500 independent Monte Carlo replications. Performance was summarized by AUC, sensitivity, specificity, PPV, and accuracy. We did not use either a train/test split or cross-validation: in other words, PheNorm models were trained and evaluated on all observations. AUC was treated as the primary threshold-free discrimination metric. Threshold-based metrics used the median score threshold, equivalently the 50th percentile of the predicted score, in non-rare settings. In rare-outcome settings, threshold-based metrics used the 0.95 score quantile threshold. This high-score threshold was chosen to reflect the practical goal of identifying a high-probability subset rather than classifying half of the cohort as positive, following the use of high-probability phenotyping thresholds in EHR-based studies \citep{coley2021clinical}.

\begin{table}[htbp]
\centering
\footnotesize
\begin{threeparttable}
\caption{Simulation performance across data-generating scenarios.}
\label{tab:simulation_results}

\setlength{\tabcolsep}{1.8pt}
\renewcommand{\arraystretch}{1.12}

\begin{tabular}{@{}llccccc@{}}
\toprule
Scenario & Silver-label setting & AUC & Sens. & Spec. & PPV & Acc. \tabularnewline
\midrule
\multirow{4}{*}{One-sided}
& 1 binary & 0.913 (0.024) & 0.827 (0.030) & 0.828 (0.031) & 0.827 (0.034) & 0.827 (0.029) \tabularnewline
& 2 binary & 0.945 (0.015) & 0.885 (0.018) & 0.885 (0.025) & 0.885 (0.028) & 0.885 (0.019) \tabularnewline
& 1 count & 0.875 (0.016) & 0.794 (0.021) & 0.794 (0.020) & 0.793 (0.023) & 0.793 (0.018) \tabularnewline
& 1 binary + 1 count & 0.955 (0.018) & 0.882 (0.028) & 0.882 (0.032) & 0.882 (0.034) & 0.882 (0.027) \tabularnewline
\midrule
\multirow{4}{*}{Two-sided}
& 1 binary & 0.908 (0.015) & 0.809 (0.012) & 0.810 (0.012) & 0.809 (0.017) & 0.809 (0.006) \tabularnewline
& 2 binary & 0.940 (0.011) & 0.867 (0.022) & 0.867 (0.023) & 0.866 (0.025) & 0.867 (0.019) \tabularnewline
& 1 count & 0.875 (0.016) & 0.794 (0.021) & 0.794 (0.020) & 0.794 (0.024) & 0.794 (0.018) \tabularnewline
& 1 binary + 1 count & 0.936 (0.009) & 0.813 (0.014) & 0.813 (0.014) & 0.813 (0.018) & 0.813 (0.010) \tabularnewline
\midrule
\multirow{4}{*}{One-sided rare}
& 1 binary & 0.873 (0.054) & 0.638 (0.070) & 0.981 (0.006) & 0.641 (0.111) & 0.964 (0.007) \tabularnewline
& 2 binary & 0.930 (0.035) & 0.840 (0.052) & 0.991 (0.006) & 0.836 (0.112) & 0.984 (0.006) \tabularnewline
& 1 count & 0.882 (0.027) & 0.427 (0.064) & 0.969 (0.004) & 0.427 (0.077) & 0.943 (0.007) \tabularnewline
& 1 binary + 1 count & 0.953 (0.026) & 0.733 (0.070) & 0.986 (0.005) & 0.728 (0.102) & 0.972 (0.007) \tabularnewline
\midrule
\multirow{4}{*}{Two-sided rare}
& 1 binary & 0.837 (0.041) & 0.267 (0.092) & 0.961 (0.005) & 0.266 (0.099) & 0.926 (0.010) \tabularnewline
& 2 binary & 0.910 (0.026) & 0.497 (0.074) & 0.973 (0.004) & 0.495 (0.085) & 0.949 (0.007) \tabularnewline
& 1 count & 0.882 (0.027) & 0.427 (0.066) & 0.970 (0.004) & 0.425 (0.076) & 0.942 (0.007) \tabularnewline
& 1 binary + 1 count & 0.879 (0.038) & 0.393 (0.114) & 0.968 (0.006) & 0.392 (0.120) & 0.939 (0.012) \tabularnewline
\bottomrule
\end{tabular}

\begin{tablenotes}[flushleft]
\footnotesize
\item[] Values are reported as mean (Monte Carlo standard deviation) over simulation replications. The silver-label setting column describes the silver labels used in the fitted model. Sensitivity, specificity, PPV, and accuracy were evaluated using the median score threshold for non-rare scenarios and the 0.95 score quantile threshold for rare-outcome scenarios. Sens. = sensitivity; Spec. = specificity; PPV = positive predictive value; Acc. = accuracy.
\end{tablenotes}
\end{threeparttable}
\end{table}

Binary PheNorm performed well. In the non-rare settings, one binary silver label gave mean AUCs of 0.913 under one-sided misclassification and 0.908 under two-sided misclassification (Table~\ref{tab:simulation_results}). Since the gold-standard phenotype was not used in model fitting, these results show that an informative binary surrogate, together with covariates, can support weakly supervised phenotype scoring.

Performance improved when a second binary silver label was available. In the non-rare settings, adding a second binary label increased mean AUC from 0.913 to 0.945 under one-sided misclassification and from 0.908 to 0.940 under two-sided misclassification. The same pattern appeared in the rare-outcome settings: mean AUC increased from 0.873 to 0.930 under one-sided misclassification and from 0.837 to 0.910 under two-sided misclassification (Table~\ref{tab:simulation_results}). These comparisons support the claim that Binary PheNorm can combine information across multiple noisy binary surrogates.

Binary silver labels also complemented count silver labels in most scenarios. In the one-sided non-rare setting, Original PheNorm using one count silver label had mean AUC 0.875, whereas the model using one binary and one count label had mean AUC 0.955. In the two-sided non-rare setting, the corresponding AUCs were 0.875 and 0.936. In the one-sided rare-outcome setting, adding a binary label to the count label increased mean AUC from 0.882 to 0.953. The exception was the two-sided rare-outcome setting, where the combined model had mean AUC 0.879, similar to Original PheNorm using one count silver label, with mean AUC 0.882. Thus, the simulations suggest that an informative binary silver label can improve Original PheNorm in many settings. Even when the binary label did not improve performance, as in the two-sided rare-outcome setting, adding it did not substantially reduce discrimination.

The threshold-based results are driven in part by the prespecified cutoffs. In the non-rare settings, the median score cutoff classified roughly half of the subjects as positive, leading to balanced sensitivity and specificity. In the rare-outcome settings, the 95th percentile cutoff selected only the highest-scoring 5\% of subjects, which produced high specificity and accuracy but lower sensitivity.

\subsection{Testing Binary PheNorm in a sparse high-dimensional setting}\label{subsec:sim_highdim}

We used a different data-generating model with a sparse signal structure to evaluate the lasso extension in a non-rare outcome setting. Each simulated dataset contained \(n=500\) subjects. The true phenotype was generated as \(y_i\sim\mathrm{Bernoulli}(0.5)\). We generated \(d=100\) covariates, of which only the first five were informative: for \(j=1,\ldots,5\), \(z_{ij}\mid y_i=1\sim N(0.6,1)\) and \(z_{ij}\mid y_i=0\sim N(0,1)\); for \(j=6,\ldots,100\), \(z_{ij}\sim N(0,1)\).

This design created a high-dimensional setting with a sparse phenotype signal, allowing us to assess whether lasso regularization improved the denoising step by reducing the influence of noise covariates.

The high-dimensional simulation generated two binary silver labels and two count silver labels. Binary silver labels satisfied \(P(b_{ij}=y_i)=0.8\). For count silver labels, we generated \(n_i\sim\mathrm{Poisson}(2)+1\) and defined latent means
\[
\eta_{i,\mathrm{ICD}}
=
\mu_{y_i}+0.8z_{i1}+0.6z_{i2}-0.5z_{i3},
\qquad
\text{and}
\qquad
\eta_{i,\mathrm{NLP}}
=
\mu_{y_i}+0.7z_{i1}+0.5z_{i4}-0.4z_{i5}.
\]
where \(\mu_0=0.5\), \(\mu_1=1.0\), and \(\sigma_0=\sigma_1=0.6\). The observed count silver labels were generated as
\[
c_{i,\mathrm{ICD}}
=
\log\left\lceil
\exp\{e_{i,\mathrm{ICD}}\}n_i^{1/2}
\right\rceil,
\qquad
\text{where}\qquad
e_{i,\mathrm{ICD}}\sim N(\eta_{i,\mathrm{ICD}},\sigma_{y_i}^2).
\]
and analogously for \(c_{i,\mathrm{NLP}}\).

The fitted models used one binary silver label and one count silver label. We compared an ordinary least-squares denoising step with a 10-fold cross-validated lasso denoising step, using the same simulated datasets in each Monte Carlo replication. The comparison was summarized over 1000 independent replications. Performance was evaluated by AUC, sensitivity, specificity, PPV, and accuracy. Because this was a non-rare outcome setting, threshold-based metrics were computed using the median score threshold.

\begin{table}[htbp]
\centering
\small
\begin{threeparttable}
\caption{Simulation performance in the sparse high-dimensional setting.}
\label{tab:sim_lasso_comparison}

\setlength{\tabcolsep}{4pt}
\renewcommand{\arraystretch}{1.15}

\begin{tabular}{@{}lccccc@{}}
\toprule
Method & AUC & Sens. & Spec. & PPV & Acc. \\
\midrule
LM PheNorm    & 0.867 (0.012) & 0.803 (0.017) & 0.802 (0.016) & 0.801 (0.024) & 0.802 (0.010) \\
Lasso PheNorm & 0.878 (0.010) & 0.808 (0.016) & 0.807 (0.017) & 0.806 (0.025) & 0.807 (0.009) \\
\bottomrule
\end{tabular}

\begin{tablenotes}[flushleft]
\footnotesize
\item[] Values are reported as mean (Monte Carlo standard deviation) over simulation replications. The sparse high-dimensional setting used one binary silver label and one count silver label. Sensitivity, specificity, PPV, and accuracy were evaluated using the median score threshold. Sens. = sensitivity; Spec. = specificity; PPV = positive predictive value; Acc. = accuracy; LM PheNorm = standard PheNorm using linear regression denoising. Lasso PheNorm = PheNorm using lasso regression denoising.
\end{tablenotes}
\end{threeparttable}
\end{table}

In the sparse high-dimensional simulation, lasso denoising produced modest but consistent improvements over ordinary least squares. Mean AUC increased from 0.867 to 0.878, and accuracy increased from 0.802 to 0.807 (Table~\ref{tab:sim_lasso_comparison}). Sensitivity, specificity, and PPV also increased slightly. Because the data-generating mechanism included only five informative covariates among 100 candidates, this result supports the use of lasso denoising when the EHR feature matrix is sparse and high-dimensional.
\section{Computable phenotyping models for anaphylaxis and acute pancreatitis}\label{sec:real_data_analysis}

\subsection{Data sources and silver labels}

We evaluated Binary PheNorm in two EHR phenotyping applications: anaphylaxis at Kaiser Permanente Washington (KPWA) and acute pancreatitis at Kaiser Permanente Northwest (KPNW). In both applications, gold-standard labels were obtained by medical record review and used for performance evaluation. The PheNorm models were fit using silver labels, structured EHR variables, and NLP-derived features. The gold-standard labels were not used to estimate the weakly supervised scores.

The anaphylaxis dataset consisted of potential anaphylaxis encounters in the KPWA EHR. The cohort was restricted to encounters from January 1, 2016 through March 31, 2023 among patients aged 16 years or older. Potential events were identified using ICD-10-CM diagnosis-code and encounter-based rules, including inpatient or emergency department encounters with an anaphylaxis diagnosis code, outpatient encounters with an anaphylaxis diagnosis code plus same-day evidence of symptoms or treatment, and inpatient or emergency department encounters with selected allergy or adverse-effect diagnosis codes plus same-day supporting codes \citep{walsh2013validation, carrell2023improving}. The full analytic dataset included 1,028 potential encounters \citep{williamson2026identifying}. Among these, 145 encounters had gold-standard labels from chart review, of which 58 were positive for anaphylaxis. The chart-reviewed sample was obtained using a stratified sampling design, and sampling weights were used for performance evaluation.

For anaphylaxis, the gold-standard outcome was the chart-review label indicating whether the encounter met anaphylaxis criteria. Count silver labels included counts of anaphylaxis-coded encounters, text mentions of anaphylaxis, notes containing UMLS/MedDRA concepts related to anaphylaxis or anaphylaxis treatment, and mentions of anaphylaxis or epinephrine \citep{bodenreider2004unified,brown1999medical}. We also evaluated two binary silver labels. The first was a structured indicator of epinephrine administration during the encounter. The second was an NLP-derived epinephrine-mention indicator, defined from the difference between the count of anaphylaxis-or-epinephrine mentions and the count of anaphylaxis mentions. These binary labels were included because epinephrine is the first-line treatment for anaphylaxis, suggesting possible utility as a silver label \citep{cardona2020world,golden2024anaphylaxis}.

The acute pancreatitis dataset consisted of potential incident acute pancreatitis events from the KPNW EHR between October 1, 2015 and December 31, 2019 \citep{bann2026comparison}. Eligible individuals were at least 18 years old, continuously enrolled for at least one year before the qualifying encounter, and had an inpatient, emergency department, or outpatient encounter at a KPNW-owned facility with an acute pancreatitis diagnosis code. Individuals with evidence of acute or chronic pancreatitis in the prior year were excluded to focus on incident events. Among 1,843 eligible potential events, a simple random sample of 386 events was selected for medical record review to obtain gold-standard labels, of which 179 were positive for acute pancreatitis. The data are described more fully in \citet{floyd2023validation}.

For acute pancreatitis, the gold-standard outcome was the chart-review label for acute pancreatitis. Count silver labels included the number of calendar days with acute pancreatitis diagnosis-coded encounters, the number of affirmative NLP mentions of acute pancreatitis, a combined diagnosis-plus-NLP silver label, and the maximum lipase laboratory value normalized to the upper limit of normal within the laboratory catchment window. The binary silver label was an indicator that the maximum lipase value exceeded three times the upper limit of normal \citep{banks2013classification, floyd2023validation}. This threshold-based binary label is clinically interpretable and routinely available.

\subsection{Analysis strategy}

For each application, we evaluated binary silver-label rules (i.e., using only the binary silver label), Original PheNorm models, Binary PheNorm models, and Combined PheNorm using both binary and count silver labels. Because the real-data feature matrices contained many sparse structured and NLP-derived predictors, all PheNorm models in the real-data analyses used lasso-regularized denoising. Therefore, the real-data comparisons evaluate silver-label configurations within a common lasso-based PheNorm framework rather than comparing ordinary least squares with lasso.

We report results from model training and evaluation using a train/test split, cross-validation, and no splitting. The train/test and cross-validation analyses evaluate possible overfitting by holding out some gold-labeled observations from model training and using them for evaluation. The no-split analysis fits the weakly supervised model using all available silver-label and covariate information and evaluates performance on the chart-reviewed sample. Because gold-standard labels are not used in model fitting, sample splitting is not required for training the weakly supervised score. The no-split analysis therefore provides the closest analogue to the final model that would be used in practice, although it is less comparable to standard supervised prediction model evaluation. Agreement across the three strategies would suggest that the results are not subject to optimism when the model is trained and evaluated on the chart-reviewed sample.

\subsection{Anaphylaxis results}

\begin{table}[htbp]
\centering
\small
\begin{threeparttable}
\caption{Predictive performance for anaphylaxis phenotyping.}
\label{tab:ana_model_performance}

\setlength{\tabcolsep}{3.2pt}
\renewcommand{\arraystretch}{1.15}

\begin{tabular}{@{}llccccccc@{}}
\toprule
Analysis & Model / silver-label set & AUC & Cutoff & Acc. & Sens. & Spec. & PPV & NPV \\
\midrule

\multicolumn{9}{@{}l}{\textit{Binary silver-label rules without PheNorm}} \\
No PheNorm & Epi admin. & 0.601 & --- & 0.572 & 0.741 & 0.460 & 0.478 & 0.727 \\
No PheNorm & Epi mention & 0.793 & --- & 0.766 & 0.931 & 0.655 & 0.643 & 0.934 \\
\addlinespace[0.45em]

\multicolumn{9}{@{}l}{\textit{Lasso-regularized PheNorm models}} \\
Train/test & Original PheNorm & 0.866 & 0.642 & 0.800 & 0.793 & 0.805 & 0.730 & 0.854 \\
& Binary PheNorm (Epi admin.) & 0.723 & 0.730 & 0.641 & 0.517 & 0.724 & 0.556 & 0.692 \\
& Binary PheNorm (Epi mention) & 0.891 & 0.715 & 0.828 & 0.776 & 0.862 & 0.789 & 0.852 \\
& Original + Binary (Epi mention) & 0.878 & 0.667 & 0.800 & 0.810 & 0.793 & 0.723 & 0.863 \\
& All labels & 0.884 & 0.671 & 0.814 & 0.793 & 0.828 & 0.754 & 0.857 \\
\addlinespace[0.35em]

CV & Original PheNorm & 0.866 & 0.644 & 0.798 & 0.789 & 0.803 & 0.728 & 0.852 \\
& Binary PheNorm (Epi admin.) & 0.713 & 0.727 & 0.606 & 0.506 & 0.673 & 0.515 & 0.670 \\
& Binary PheNorm (Epi mention) & 0.891 & 0.717 & 0.826 & 0.760 & 0.870 & 0.796 & 0.846 \\
& Original + Binary (Epi mention) & 0.880 & 0.657 & 0.798 & 0.795 & 0.801 & 0.727 & 0.855 \\
& All labels & 0.885 & 0.659 & 0.815 & 0.808 & 0.820 & 0.750 & 0.865 \\
\addlinespace[0.35em]

No split & Original PheNorm & 0.867 & 0.593 & 0.793 & 0.776 & 0.805 & 0.726 & 0.843 \\
& Binary PheNorm (Epi admin.) & 0.758 & 0.723 & 0.724 & 0.603 & 0.805 & 0.673 & 0.753 \\
& Binary PheNorm (Epi mention) & 0.892 & 0.716 & 0.828 & 0.776 & 0.862 & 0.789 & 0.852 \\
& Original + Binary (Epi mention) & 0.878 & 0.665 & 0.793 & 0.793 & 0.793 & 0.719 & 0.852 \\
& All labels & 0.883 & 0.669 & 0.821 & 0.810 & 0.828 & 0.758 & 0.867 \\
\bottomrule
\end{tabular}

\begin{tablenotes}[flushleft]
\footnotesize
\item Values are weighted estimates evaluated on chart-reviewed encounters. All PheNorm models used lasso-regularized denoising. Original PheNorm denotes the model using count silver labels. Epi admin. denotes the structured epinephrine administration indicator, and Epi mention denotes the NLP-derived epinephrine-mention indicator. All labels denotes the model using count silver labels together with both binary silver labels. Cutoff is the 60th percentile of the predicted score, used to compute threshold-based metrics. The rows without PheNorm evaluate binary silver-label rules directly. AUC = area under the receiver operating characteristic curve; Acc. = accuracy; Sens. = sensitivity; Spec. = specificity; PPV = positive predictive value; NPV = negative predictive value; CV = cross-validation.
\end{tablenotes}
\end{threeparttable}
\end{table}

The epinephrine-mention binary silver label was informative even before applying PheNorm, with AUC 0.793 as a raw binary rule (Table~\ref{tab:ana_model_performance}). Binary PheNorm improved discrimination substantially. Using the epinephrine-mention label alone, the AUC was 0.891 in the train/test analysis, 0.891 in cross-validation, and 0.892 in the no-split analysis. The Binary PheNorm model also produced higher specificity and PPV than the raw epinephrine-mention rule, although sensitivity was lower. This pattern suggests that the denoising step reduced false positives and produced a more specific phenotype score. These results support the first empirical motivation for Binary PheNorm: a binary silver label can be useful by itself, and the PheNorm denoising step can improve over the raw binary rule.

The epinephrine-mention Binary PheNorm model also outperformed the Original PheNorm model in this application. The Original PheNorm model had AUCs of 0.866 in the train/test analysis, 0.866 in cross-validation, and 0.867 in the no-split analysis, whereas the binary epinephrine-mention model had AUCs of 0.891, 0.891, and 0.892, respectively. Adding the epinephrine-mention binary label to Original PheNorm improved AUC relative to Original PheNorm alone: from 0.866 to 0.878 in the train/test analysis, from 0.866 to 0.880 in cross-validation, and from 0.867 to 0.878 in the no-split analysis. These comparisons indicate that the binary epinephrine-mention label contained phenotype information not fully captured by the Original PheNorm model.

The structured epinephrine administration indicator was less informative than the epinephrine-mention indicator. Its raw AUC was 0.601, and the Binary PheNorm AUCs ranged from 0.713 to 0.758. This result is important because it shows that Binary PheNorm does not make an uninformative binary label useful by assumption. Its performance depends on the clinical relevance and data quality of the binary surrogate.

We also repeated the single-binary-label Binary PheNorm analyses using 20 additional random seeds to assess numerical stability. The Binary PheNorm model based on the structured epinephrine administration indicator had AUCs ranging from 0.690 to 0.758, whereas the model based on the epinephrine-mention indicator had AUCs ranging from 0.882 to 0.897. This additional check suggests that the more informative binary silver label also produced more stable discrimination across random seeds.

\subsection{Acute pancreatitis results}

\begin{table}[htbp]
\centering
\small
\begin{threeparttable}
\caption{Predictive performance for acute pancreatitis phenotyping.}
\label{tab:ap_model_performance}

\setlength{\tabcolsep}{3.2pt}
\renewcommand{\arraystretch}{1.15}

\begin{tabular}{@{}llccccccc@{}}
\toprule
Analysis & Model / silver-label set & AUC & Cutoff & Acc. & Sens. & Spec. & PPV & NPV \\
\midrule

\multicolumn{9}{@{}l}{\textit{Binary silver-label rule without PheNorm}} \\
No PheNorm & Lipase $>3\times$ ULN & 0.736 & --- & 0.686 & 0.497 & 0.975 & 0.968 & 0.558 \\
\addlinespace[0.45em]

\multicolumn{9}{@{}l}{\textit{Lasso-regularized PheNorm models}} \\
Train/test & Original PheNorm & 0.857 & 0.538 & 0.720 & 0.598 & 0.925 & 0.930 & 0.579 \\
& Original PheNorm (no lab) & 0.745 & 0.546 & 0.654 & 0.559 & 0.813 & 0.833 & 0.524 \\
& Binary PheNorm & 0.805 & 0.508 & 0.710 & 0.587 & 0.916 & 0.921 & 0.570 \\
& All labels & 0.883 & 0.536 & 0.720 & 0.587 & 0.944 & 0.946 & 0.577 \\
\addlinespace[0.35em]

CV & Original PheNorm & 0.855 & 0.550 & 0.717 & 0.592 & 0.926 & 0.931 & 0.576 \\
& Original PheNorm (no lab) & 0.754 & 0.550 & 0.650 & 0.549 & 0.819 & 0.836 & 0.521 \\
& Binary PheNorm & 0.809 & 0.513 & 0.721 & 0.593 & 0.935 & 0.938 & 0.579 \\
& All labels & 0.888 & 0.562 & 0.716 & 0.589 & 0.929 & 0.933 & 0.575 \\
\addlinespace[0.35em]

No split & Original PheNorm & 0.857 & 0.541 & 0.717 & 0.598 & 0.916 & 0.922 & 0.576 \\
& Original PheNorm (no lab) & 0.750 & 0.543 & 0.650 & 0.547 & 0.822 & 0.838 & 0.521 \\
& Binary PheNorm & 0.819 & 0.511 & 0.717 & 0.592 & 0.925 & 0.930 & 0.576 \\
& All labels & 0.891 & 0.573 & 0.713 & 0.587 & 0.925 & 0.929 & 0.572 \\
\bottomrule
\end{tabular}

\begin{tablenotes}[flushleft]
\footnotesize
\item Values are weighted estimates evaluated on chart-reviewed events. All PheNorm models used lasso-regularized denoising. Original PheNorm denotes the model using count/non-binary silver labels, and Original PheNorm (no lab) excludes the normalized lipase laboratory value. All labels denotes the model using all available silver labels. Cutoff is the 60th percentile of the predicted score used to compute threshold-based metrics. The row without PheNorm evaluates the binary laboratory rule directly. AUC = area under the receiver operating characteristic curve; Acc. = accuracy; Sens. = sensitivity; Spec. = specificity; PPV = positive predictive value; NPV = negative predictive value; ULN = upper limit of normal; CV = cross-validation.
\end{tablenotes}
\end{threeparttable}
\end{table}

The binary lipase-threshold rule was informative as a raw silver-label rule, with AUC 0.736 (Table~\ref{tab:ap_model_performance}). Binary PheNorm improved discrimination relative to this raw rule, with AUC 0.805 in the train/test analysis, 0.809 in cross-validation, and 0.819 in the no-split analysis. This comparison is important because both approaches use the same binary laboratory indicator, but Binary PheNorm additionally uses the denoising regression to borrow information from auxiliary EHR and NLP-derived covariates. The improvement therefore supports the role of the PheNorm denoising step for this binary silver label.

The Original PheNorm model achieved higher AUCs than the binary-only model, with AUCs 0.857, 0.855, and 0.857 across the train/test, cross-validation, and no-split analyses. However, this comparison should be interpreted carefully because the Original PheNorm model included the continuous normalized lipase laboratory value, which contains information closely related to the binary lipase-threshold label \citep{bann2026comparison}. For this reason, the more direct comparison for evaluating the usefulness of the binary laboratory label is between Binary PheNorm and Original PheNorm excluding the laboratory value. In that comparison, Binary PheNorm performed better across all three evaluation strategies: AUC increased from 0.745 to 0.805 in the train/test analysis, from 0.754 to 0.809 in cross-validation, and from 0.750 to 0.819 in the no-split analysis. These results indicate that the binary lipase-threshold label carried substantial phenotype information and that Binary PheNorm was able to use this information effectively.

The Combined PheNorm produced the best discrimination in all three evaluation strategies, with AUC 0.883 in the train/test analysis, 0.888 in cross-validation, and 0.891 in the no-split analysis. Compared with Original PheNorm, adding the binary lipase-threshold label increased AUC from 0.857 to 0.883 in the train/test analysis, from 0.855 to 0.888 in cross-validation, and from 0.857 to 0.891 in the no-split analysis. This improvement suggests that the binary threshold captured clinically meaningful information that was not fully represented by the Original PheNorm inputs, including the continuous normalized laboratory value and the other count-based silver labels. Under the train/test split, the all-label model also increased specificity from 0.925 to 0.944 and PPV from 0.930 to 0.946 relative to Original PheNorm, indicating that the combined model reduced false positives at the prespecified threshold.

The three evaluation strategies gave similar AUCs in both applications. For anaphylaxis, the epinephrine-mention Binary PheNorm model had AUCs 0.891, 0.891, and 0.892. For acute pancreatitis, the all-label model had AUCs 0.883, 0.888, and 0.891. These similarities support the interpretation that the weakly supervised scores were not highly sensitive to the particular evaluation split. They also support the practical use of fitting the final model on the full available dataset when the goal is to construct a phenotype score using all silver-label information.

\section{Discussion}\label{sec:discussion}

In this study, Binary PheNorm and its lasso-regularized extension showed three main findings. First, informative binary silver labels can be used directly within the PheNorm denoising framework to construct weakly supervised phenotype scores, and this denoising step improved substantially over using the binary indicators alone in both real-data applications. Second, lasso-regularized denoising provided modest but consistent gains in sparse high-dimensional simulations and offered a practical implementation for real EHR feature matrices with many sparse structured and NLP-derived predictors. Third, Binary PheNorm can be integrated into the broader PheNorm framework, allowing binary silver labels to be used alongside count or continuous silver labels when these sources contain complementary phenotype information.

The first contribution is that Binary PheNorm provides a weakly supervised denoising approach for binary silver labels. It preserves the main structure of PheNorm because the score is estimated from silver labels and covariates, while gold-standard labels are used only for evaluation. This distinction is important because chart review is often the limiting resource in EHR phenotyping studies. Binary PheNorm also avoids imposing the mixture-normal assumption used by Original PheNorm, which is appropriate because a binary silver label is not naturally modeled through the same transformed count-label mixture structure. The empirical results support this finding. In anaphylaxis, the raw epinephrine-mention rule had AUC 0.793, whereas Binary PheNorm using the same binary label had AUCs 0.891, 0.891, and 0.892 across the train/test, cross-validation, and no-split analyses. In acute pancreatitis, the raw laboratory-threshold rule had AUC 0.736, whereas Binary PheNorm improved AUC to 0.805 in the train/test analysis and 0.819 in both cross-validation and no-split analyses. These comparisons are informative because they compare each raw binary rule with a denoised score built from the same binary label and covariates. At the same time, the results show that binary labels are not automatically superior to count or continuous surrogates. The structured epinephrine administration label was much less informative than the epinephrine-mention label, and in acute pancreatitis the binary laboratory label alone did not outperform the Original PheNorm model. Thus, the usefulness of Binary PheNorm depends on the clinical specificity, measurement quality, and information overlap of the binary surrogate.

The second contribution is that lasso-regularized denoising can improve stability in sparse high-dimensional settings. This extension is needed because EHR and NLP feature matrices are often sparse, high-dimensional, and redundant, so an unregularized denoising regression can be unstable or overly influenced by noise variables. Lasso regularization addresses this problem by shrinking weak predictors and selecting a smaller set of informative features. In the sparse high-dimensional simulation, where only five of 100 covariates carried signal, lasso denoising improved AUC from 0.867 to 0.878 and slightly improved the threshold-based metrics. This improvement is modest but consistent with the simulation design, because regularization should be most helpful when many candidate predictors are uninformative. The same reasoning motivated the use of lasso in the real-data analyses, where the candidate predictors included many sparse structured and NLP-derived variables. Importantly, this extension is not limited to Binary PheNorm; it can also be applied to Original PheNorm and to combined models using both binary and count silver labels.

The third contribution is that Binary PheNorm can be integrated into the broader PheNorm framework described in Section~\ref{sec:methods}. Original PheNorm remains appropriate when informative count or continuous silver labels are available and its distributional assumptions are reasonable. Binary PheNorm extends this framework to settings where clinically meaningful binary silver labels are available and mixture-normal count-label calibration is not well justified. Combined PheNorm further allows binary and count or continuous silver labels to be used together when they contain complementary information. This integration is a practical contribution because many EHR phenotyping applications include heterogeneous silver labels, such as diagnosis-code counts, NLP mention counts, medication indicators, and laboratory-threshold indicators. In the acute pancreatitis application, the all-label model had the highest AUC across all three evaluation strategies, illustrating the potential benefit of combining different sources of silver-label information.

Several limitations should be noted. Binary PheNorm cannot create phenotype information from an uninformative binary surrogate; its performance depends on the clinical relevance and measurement quality of the binary silver label and the auxiliary covariates. In the anaphylaxis application, for example, the structured epinephrine administration indicator was much less informative than the epinephrine-mention indicator. A second limitation is that the Binary PheNorm output is a phenotype score rather than a calibrated phenotype probability, because the linear denoising regression is not constrained to produce calibrated probabilities. If probability-scale outputs are required, truncation to \([0,1]\), an inverse-logit transformation, or calibration using an independent gold-standard validation sample would be needed. Further evaluation in additional outcomes and health systems is needed because the current real-data evidence comes from two clinical phenotyping applications.

\section*{Code availability}

The R code implementing Binary PheNorm and Combined PheNorm, together with code to reproduce the simulation studies, is publicly available at \url{https://github.com/Shuhe-W/Binary-PheNorm}. The analysis code for the electronic health record applications and the underlying restricted data are not publicly available because of institutional and data-use restrictions; they may be made available upon reasonable request and subject to the necessary approvals.

\bibliographystyle{chicago}
\bibliography{references}

\end{document}


\maketitle

\section{Comparison of logistic-regression and linear-regression denoising}
\label{app:logistic_binary_phenorm}

In the main Binary PheNorm implementation, the binary silver label is denoised using ordinary least-squares regression rather than logistic regression. Although logistic regression is a natural model for a binary response, the response in the denoising step is the silver label rather than the gold-standard phenotype. The purpose of the denoising step is therefore to construct a stable continuous phenotype score, rather than to estimate a calibrated conditional probability of the true phenotype.

We conducted a supplementary simulation experiment comparing logistic-regression and linear-regression denoising under the non-rare, two-sided binary silver-label data-generating scenario used in the main simulations. In each Monte Carlo replication, we generated \(n=1000\) observations with
\[
Y_i \sim \mathrm{Bernoulli}(0.5).
\]
Ten auxiliary covariates were generated from phenotype-specific normal distributions,
\[
Z_{ij}\mid Y_i=1 \sim N(0.7,0.5^2),
\qquad
Z_{ij}\mid Y_i=0 \sim N(0.3,0.5^2),
\qquad j=1,\ldots,10.
\]
Two binary silver labels were generated by independently flipping 20\% of the true phenotype labels, so that each binary silver label had 80\% agreement with the true phenotype. This corresponds to the two-sided misclassification setting in the main simulation design.

For each simulated dataset, we compared four scores: the raw single binary silver label, the average of two raw binary silver labels, Binary PheNorm using logistic-regression denoising, and Binary PheNorm using linear-regression denoising. The logistic- and linear-regression Binary PheNorm scores used the same single binary silver label, the same auxiliary covariates, and the same corruption rate of 0.4. Thus, the comparison isolates the effect of replacing the linear denoising regression with a logistic denoising regression. The simulation was repeated over 1000 independent Monte Carlo replications, and performance was summarized using the area under the receiver operating characteristic curve (AUC).

\begin{table}[htbp]
\centering
\begingroup
\singlespacing
\small
\captionsetup{skip=4pt}
\begin{threeparttable}
\caption{Monte Carlo comparison of logistic-regression and linear-regression denoising in Binary PheNorm.}
\label{tab:logistic_binary_phenorm_mc}
\setlength{\tabcolsep}{5pt}
\renewcommand{\arraystretch}{1.15}
\begin{tabular}{@{}lccc@{}}
\toprule
Method & Mean AUC (SD) & Median AUC & 2.5--97.5\% range \\
\midrule
Raw single binary label & 0.800 (0.000) & 0.800 & 0.799--0.801 \\
Raw average of two binary labels & 0.896 (0.003) & 0.896 & 0.890--0.903 \\
Logistic Binary PheNorm & 0.658 (0.004) & 0.657 & 0.651--0.666 \\
Linear Binary PheNorm & 0.906 (0.015) & 0.907 & 0.876--0.934 \\
\bottomrule
\end{tabular}
\begin{tablenotes}[flushleft]
\footnotesize
\item[] Values are reported as mean (Monte Carlo standard deviation), median, and empirical 2.5th--97.5th percentile range over 1000 Monte Carlo replications. The logistic- and linear-regression Binary PheNorm rows used the same simulated datasets, single binary silver label, auxiliary covariates, and corruption mechanism; only the regression model in the denoising step was changed. AUC = area under the receiver operating characteristic curve; SD = standard deviation.
\end{tablenotes}
\end{threeparttable}
\endgroup
\end{table}

\begin{figure}[htbp]
\centering
\includegraphics[width=0.82\textwidth]{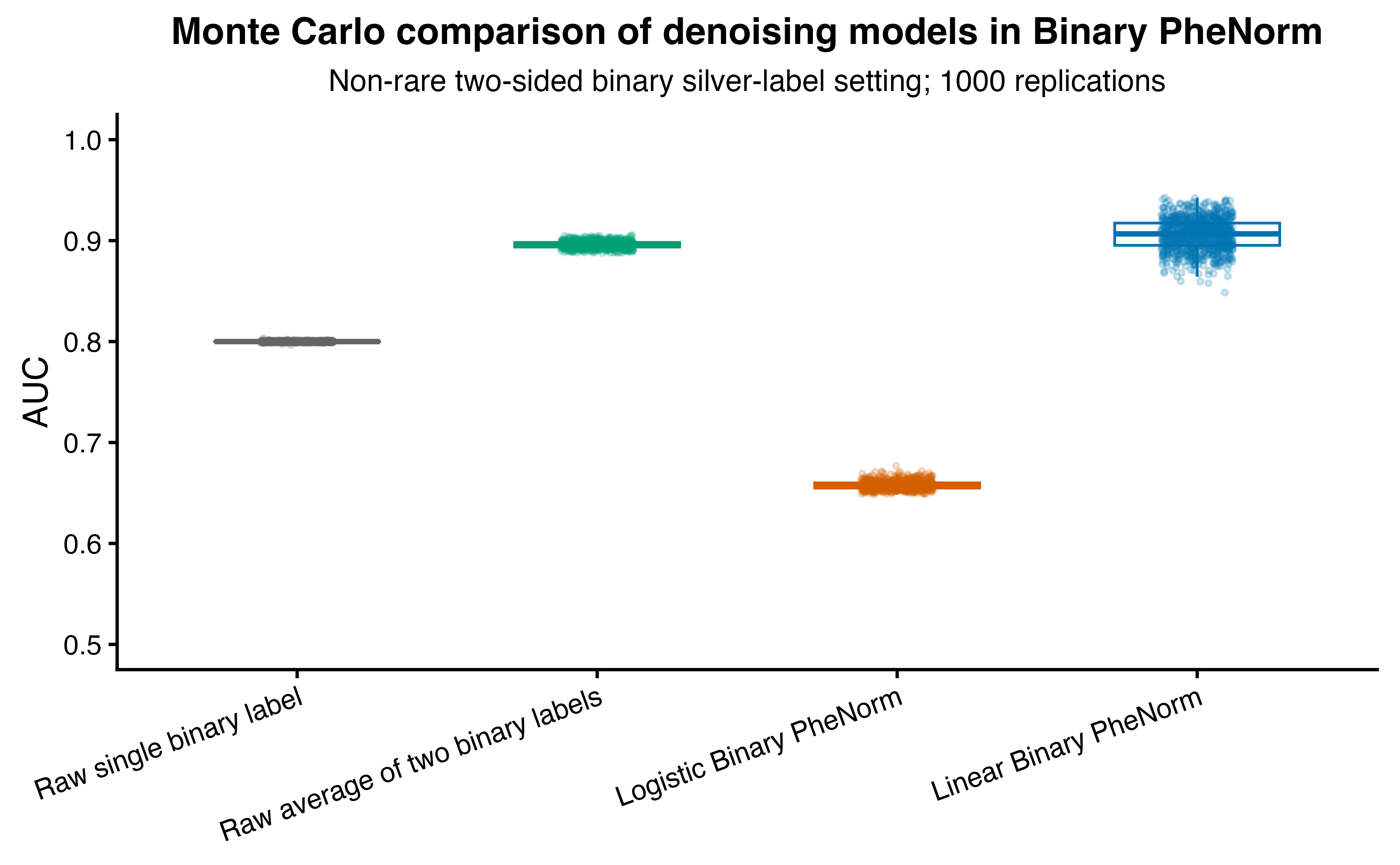}
\caption{Distribution of AUC values across 1000 Monte Carlo replications under the non-rare, two-sided binary silver-label setting. The figure compares raw binary silver-label rules, logistic-regression denoising, and linear-regression denoising in Binary PheNorm.}
\label{fig:appendix_auc_comparison}
\end{figure}

The Monte Carlo results in Table~\ref{tab:logistic_binary_phenorm_mc} and Figure~\ref{fig:appendix_auc_comparison} summarize the discrimination of each score across simulated datasets. These results compare the two regression choices for the denoising step in Binary PheNorm.

\begin{figure}[htbp]
\centering
\includegraphics[width=0.72\textwidth]{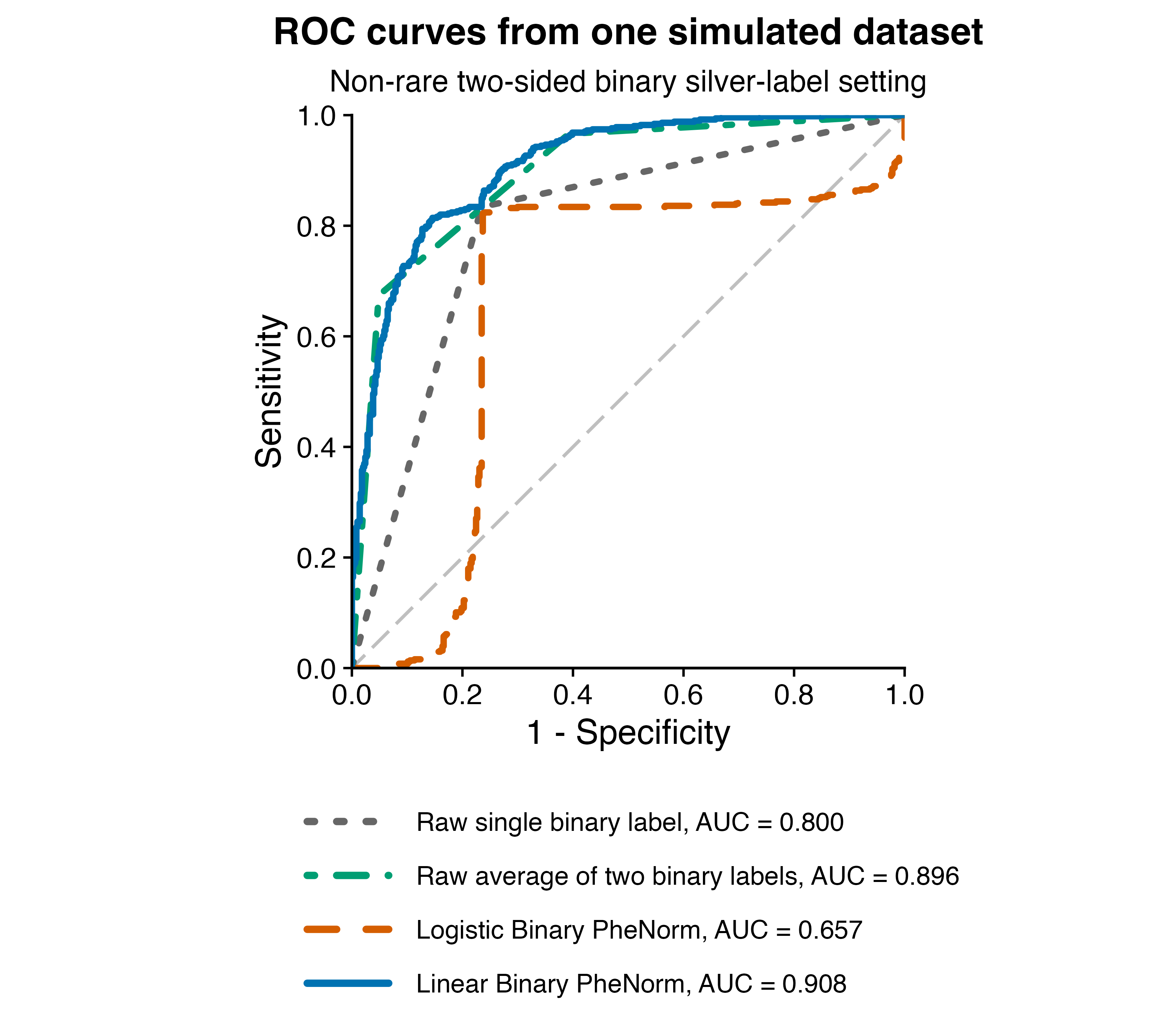}
\caption{Representative ROC curves from one simulated dataset under the non-rare, two-sided binary silver-label setting. The curves compare raw binary silver-label rules, logistic-regression denoising, and linear-regression denoising in Binary PheNorm. The AUC values shown in the legend are computed from this representative simulated dataset.}
\label{fig:appendix_sample_roc_curve}
\end{figure}

Figure~\ref{fig:appendix_sample_roc_curve} shows representative ROC curves from one simulated dataset. This figure is intended only as a visual illustration of the ranking behavior of the different scores in a single simulated sample; the Monte Carlo summaries in Table~\ref{tab:logistic_binary_phenorm_mc} and Figure~\ref{fig:appendix_auc_comparison} provide the primary simulation evidence.

Overall, this supplementary comparison evaluates whether logistic-regression denoising is appropriate for the Binary PheNorm denoising step. The results support the use of linear-regression denoising in the main Binary PheNorm analyses and reinforce the interpretation of the Binary PheNorm output as a continuous phenotype score rather than a calibrated phenotype probability.